\documentclass[prl,final,twocolumn,floatfix]{revtex4-1}

\usepackage{amsmath,amssymb}
\usepackage{graphicx}
\usepackage{times,mathptmx} 
\usepackage[pdftex,
  colorlinks=true,%
  urlcolor=blue,%
  linkcolor=blue,%
  citecolor=blue,%
  pdftitle={Why do nanotubes grow chiral?},%
  pdfauthor={Evgeni Penev}%
]{hyperref}
\usepackage[all]{hypcap}

\newcommand{\fig}{\textbf{Fig.}}

\begin{document}

\title{Why Do Nanotubes Grow Chiral?}

\author{Vasilii I. Artyukhov}\thanks{These authors contributed equally.}
\author{Evgeni S. Penev}\thanks{These authors contributed equally.}
\author{Boris I. Yakobson}\email[Correspondence to: ]{biy@rice.edu}

\affiliation{Department of Materials Science and NanoEngineering, Rice
  University, Houston, TX 77005, USA}

\begin{abstract}
Carbon nanotubes (CNT) hold enormous technological promise. It can only be
harnessed if one controls in a practical way the CNT chirality, the feature of
the tubular carbon topology that governs all the CNT properties---electronic,
optical, mechanical. Experiments in catalytic growth over the last decade have
repeatedly revealed a puzzling strong preference towards minimally-chiral
(near-armchair) CNT, challenging any existing hypotheses and turning chirality
control ever more tantalizing, yet leaving its understanding elusive. Here we
combine the CNT/catalyst interface thermodynamics with the kinetic growth
theory to show that the unusual near-armchair peaks emerge from the two
antagonistic trends: energetic preference towards achiral CNT/catalyst
interfaces vs. faster growth of chiral CNT. This narrow distribution is
profoundly related with the peaked behavior of a simple function, $x\,e^{-x}$.
\end{abstract}

\date{\today}

\maketitle

The broad interest in carbon nanotubes (CNT), unceasing since their first
clear observation~\cite{iijima91}, has been fueled by possible technological
applications derived from their unique fundamental
properties~\cite{dresselhaus96,avouris07,volder13}. All of the latter are in
turn determined by the helical fashion of folding a tube, specified by the
chiral angle $\chi$ between its circumference and the zigzag motif in the
honeycomb lattice of atoms, with $\chi=0^\circ$ and $\chi=30^\circ$ for the
achiral types, zigzag and armchair. Alternatively, a pair of chiral indices
$(n,m)$ is commonly used, the integer components of the
circumference-vector~\cite{dresselhaus96}. In spite of such defining role of
chirality, most synthetic methods yield a broad distribution with mixed
properties. To achieve control of the CNT type remains a great challenge; what
physical mechanisms determine the chirality distribution, and even why the
nanotubes grow chiral is still unsettled.

Although the discovery paper by Iijima~\cite{iijima91} has already suggested
one key, connecting the tube ability to grow with the kinks at its end and the
screw dislocation model, yet it took nearly two decades till the first
equation~\cite{ding09} related the speed of growth and chirality,
$R\sim\sin\chi$. It should further be useful to think, in hindsight, of the
probable causes of such delay. Besides the difficulties of determining
chirality in experiment, in theory it was ambiguous whether the chiral angle
must be measured from the zigzag (as $\chi$) or perhaps from the armchair (as
$\chi^- \equiv 30^\circ -\chi$) direction. While either choice appears valid
from pure symmetry standpoint, it changes the kinetic prediction to the
opposite, and thus one stumbles upon an immediate contradiction. Another
diversion was due to simple thermodynamic argument that the lower energy of
the tube edge, rather than its kinetic advantage of having kinks, must
determine the dominant CNT type, pointing towards the armchair tubes,
especially $(10,10)$ broadly discussed by Smalley~\textit{et
  al.}~\cite{thess96}.

This situation, together with recent advances in synthesis showing in several
cases very narrow chiral distributions~\cite{fouquet12,wang13,he13}, poses a
compelling question of which factors---thermodynamic preference to lower energy,
or kinetic preference of higher speed---play major role in defining the
distribution of CNT product. The true answer appears ``both'', and the
analysis below shows how the subtle interplay of these physical factors
defines the more probable chirality choices. In particular, it explains why at
lower temperature on solid catalyst particles the yield is peaked near
armchair type $(n,n-1)$, never exactly armchair, although quite close.

The evolution of chemical vapor deposition (CVD) techniques and chirality
characterization methods~\cite{bachilo02} has led to improvements in chiral
selectivity~\cite{fouquet12,wang13,he13,bachilo03,lolli06,li07,harutyunyan09,chiang09,liu12},
eventually reaching $>50$\% fraction for a single CNT type and $\sim 90$\% for
semiconducting tubes~\cite{he13}. More intuitive strategies such as
post-growth selection~\cite{zheng03,tu09}, rational
synthesis~\cite{scott12,mueller12,omachi13}, or
seeding~\cite{ogrin07,orbaek11,liu12b} do, in principle, offer great
selectivity but lack the scalability of CVD. Further improvements of the
latter direct synthesis techniques thus call for better understanding of the
growth process. Yet its mechanism and especially the causes behind its
occasional success in chiral preference remain puzzling. In modeling efforts,
direct molecular dynamics (MD) simulations, although invaluable in many
respects, generally fail to produce non-defective CNT structures of a
well-defined diameter and recognizable chirality, regardless of the precision
of interatomic potentials---from
classical~\cite{shibuta02,ding04,zhao05,ribas09} to
tight-binding~\cite{amara08,page10} and density functional theory
(DFT)~\cite{raty05}. This is caused by the short time-scale due to sheer
computational limitations. It is clear that a physical theory bridging the gap
between atomistic dynamics and macroscopic scales is needed to interpret both
the experiments and simulations.

At proper conditions, the nanotubes \emph{nucleation} occurs, with probability
$N_{n,m}$ of certain chiral type. It is followed by the steady carbon
accretion by each tube, with its \emph{growth rate} $R_{n,m}$. After some
time, the fraction of the tubes of chirality $(n,m)$, i.e., their relative
abundance $A_{n,m}$ in the accumulated material, is determined by the product
of both these factors~\cite{penev12} as $A_{n,m}=N_{n,m}\cdot R_{n,m}$. Using
instead of the chiral indexes the tube diameter $d$ and chiral angle $\chi$
one has, $A(\chi,d)=N(\chi,d)\cdot R(\chi,d)$. Below we explore the physical
mechanisms defining the right-hand side, in particular the case of solid
catalyst with rigid shape, which yields sharp chiral angle selectivity in
$A(\chi,d)$, as empirical evidence suggests.

During \textbf{nucleation}, as carbon atoms attach to a nascent CNT nucleus,
adding new hexagonal and pentagonal rings to it, the chirality of a CNT
becomes permanently ``locked in'' when the final $6^\text{th}$ pentagon is
added to the hemispherical cap. From this 6-pentagon nucleus a cylindrical CNT
structure can further grow by adding only hexagons, in a periodic fashion. The
free energy of the critical nucleus contains two contributions,
$G^*=G_\text{cap}+\Gamma$. The first one, $G_\text{cap}$, originates from the
``elastic'' energy of cap \textit{per se} and does not depend on
$\chi$~\cite{penev14}. The second term $\Gamma$ represents the contact
interface between the $sp^2$-carbon lattice edge and the metal catalyst, and
does contain chirality dependence since the edge energy $\gamma(\chi)$ varies
with the crystallographic orientation and $\Gamma(\chi,d)\equiv \pi d
\gamma(\chi)$. Whereas previous studies on edge energetics~\cite{liu10}
assumed either vacuum or a liquid-like catalyst that fully adapts to the edge
shape, chiral-selective CVD growth is usually reported at comparatively low
temperatures~\cite{wang13,he13} when the catalyst particle is
solid~\cite{he13}. Accordingly, the metal side of the interface is rather a
rigid atomic plane, and the structure of this interface affects both the
energy of the nucleus and the subsequent insertion of new C-atoms during
growth.

\begin{figure}[t]
\includegraphics[width=\columnwidth]{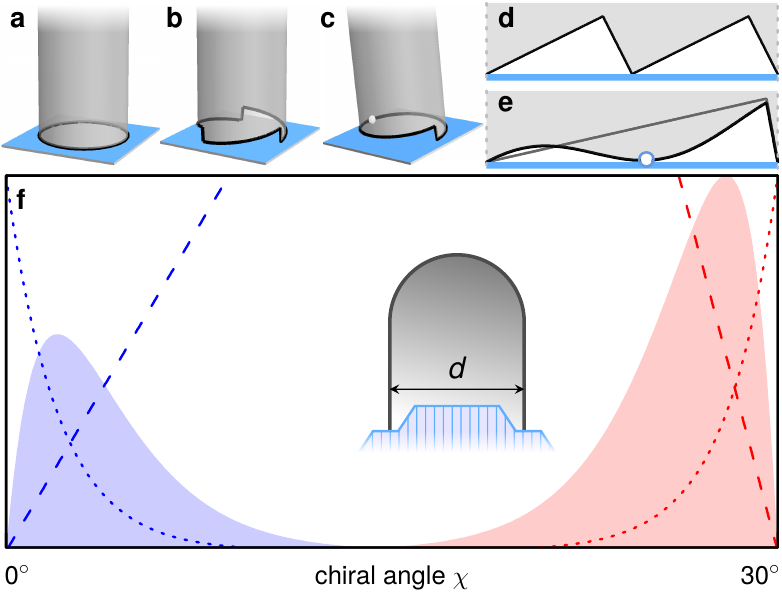}
\caption{\label{fig:1}\textbf{Continuum model of the CNT--catalyst system.} Schematic
  representation of (\textbf{a}) achiral, (\textbf{b}) multiple-kink chiral,
  and (\textbf{c}) single-kink chiral CNT on a flat substrate. Unrolled
  CNT--substrate interfaces for (\textbf{d}) two-kink and (\textbf{e})
  single-kink nanotubes show the nanotube tilt off the vertical, reducing the
  edge-substrate gap; the white dot in (\textbf{c}) and (\textbf{e}) marks the
  contact point. The abundance distributions $A(\chi) \sim \chi\, e^{-\chi}$
  computed as the product of nucleation (dotted) and growth rate (dashed)
  terms are shown in (\textbf{f}) for near-$Z$ (blue) and near-$A$ (red)
  chiralities. The inset illustrates a nascent CNT of diameter $d$ on a solid
  catalyst.}
\end{figure}

Before discussing the details of atomistic study, it is useful to explore the
key ideas in terms of simpler \textbf{continuum model}, which not only offers
valuable insight but is even able to make accurate overall predictions. In
\fig~\ref{fig:1} inset, the CNT is in contact with the catalyst which is
represented locally as a continuous plane corresponding to an atomic terrace
in a solid particle. The CNT is also continuous, but the kinks around its edge
are retained according to the tube chirality,
\fig~\ref{fig:1}\textbf{a--c}. These kinks cause the gaps between the
substrate and the CNT, shown in \fig~\ref{fig:1}\textbf{b--c} for $(n,3)$ and
$(n,1)$ tubes, with an associated energy penalty, relative to the tight
contact in case of achiral tube in \fig~\ref{fig:1}\textbf{a}. For the
hexagonal lattice of CNT the two fundamental achiral edges---armchair ($A$)
and zigzag ($Z$)-form tight low-energy contacts. The interface energy for
chiral tubes is higher, roughly in proportion with the number of kinks, which
raises linearly with $\chi$ for near-$Z$ tubes, or with $\chi^-$ for near-$A$
tubes. In other words, $\gamma(x)\approx \gamma + \gamma\,^\prime\cdot x$,
where $x$ is the angular deviation from the achiral direction: near the
$Z$-type $x=\chi$, $\gamma = \gamma_{\,Z} \equiv \gamma(0^\circ)$, and
$\gamma\,^\prime=\partial \gamma/\partial \chi |_{\chi=0^\circ}$, or near the
$A$-type $x = \chi^-$, $\gamma = \gamma_A \equiv \gamma
(30^\circ)$, and $\gamma\,^\prime = -\partial \gamma/\partial
\chi|_{\chi=30^\circ}$.

Since $\gamma(x)$ is largest in the intermediate range of
$\chi \approx 15^\circ$, such tubes are unlikely to nucleate, and one should
focus on just the neighborhoods of $Z$ and $A$ chiralities. Then we write
(omitting for brevity the $k_\text{B}T$ factor, wherever obvious):
\begin{equation}
	N(\chi,d) \propto e^{-G^*} 
        \propto e^{-\pi d (\gamma + \gamma\,^\prime \cdot x)}.
\label{eq:1}
\end{equation}
The essential result here is that the nucleation probability falls rapidly as
$e^{-\beta\cdot x}$ with chiral angle $x$ and $\beta = \pi d
\gamma\,^\prime/k_\text{B}T$. A distinction for single-kink cylinders,
representing the nanotubes $(n,1)$ and $(n,n-1)$, should be noted. Their
symmetry allows them to tilt in the vertical plane, improving the interface
contact. \fig~\ref{fig:1}\textbf{d--e} illustrates it by ``unrolling'' the
CNT--substrate interface area for two-kink and single-kink tubes. In the
latter case, the effect of tilt leads to a reduction of the tube--substrate
separation, appearing as a sinusoid along the circumference as shown in
\fig~\ref{fig:1}\textbf{e}, enabling a substantial closure of the gap between
substrate and tube edge, recovering up to as much as 70\% of the energy
penalty according to our estimates.

For the \textbf{growth rate} term $R(\chi,d)$ we augment the screw dislocation
model~\cite{ding09} by including the kinks created by thermal fluctuations on
$A$ and $Z$ edges~\cite{artyukhov12}, and accounting for the energy penalty
$\sim 1/d^2$ from the wall curvature. In the liquid-catalyst model, when the
metal adapts to the CNT edge with a one-to-one termination, calculations
suggest that the cost $E_A$ to create a pair of kinks on an $A$ edge is zero,
and consequently $R\propto \chi$~\cite{ding09}. However, on a solid surface,
creating a pair of kinks destroys the perfect contact between the CNT and
substrate, costing energy. Therefore $E_A$ has a noticeable magnitude, and the
dependence becomes bimodal with minima at the $A$ and $Z$ ends of the chiral
angle range, and a maximum at the magic angle of
$19.1^\circ$~\cite{artyukhov12,rao12}. The final expression, linearized near
the $A$ and $Z$ bounds of chirality reads as follows,
\begin{equation}
  R(\chi,d) \propto \pi d \, e^{-2C/d^2} (x+e^{-E}), 
\text{ nearly as}  \propto x,
\label{eq:2}  
\end{equation}
where $C=3.9$~eV$\cdot$\AA$^2$/atom is the bending rigidity of
graphene~\cite{kudin01}. The term $x$ in parentheses corresponds to the
density of geometry-imposed kinks, proportional to the vicinal-edge angular
deviation from the main achiral direction, and the term $e^{-E}$ (typically
small) represents the additional fluctuational kinks. The free energy barriers
for the initiation of a new atomic row on $A$ or $Z$ edge are $E = E_Z$ near
the $Z$-type where $x=\chi/\sqrt{3}$, or $E = E_A$ near the $A$-type where
$x=\chi^-$.

Multiplying together the nucleation and growth terms presents the key to
understanding the observed selectivity for near-$A$
chiralities~\cite{he13}. At a given diameter,
\begin{equation}
  A(\chi) = N(\chi)\, R(\chi) \sim \chi\, e^{-\chi},
\label{eq:3}  
\end{equation}
a function with a sharp peak near zero (or near $30^\circ$). \emph{This is the
  essential result of our continuum consideration}. \fig~\ref{fig:1}\textbf{f}
illustrates this peaked distribution character. The two distributions for
near-$Z$ (blue) and near-$A$ (red) chiral angles are plotted assuming equal
interface energies and growth barriers: $\gamma_A = \gamma_{\,Z}$, $E_A =
E_Z$. However, if either $A$ or $Z$ has a lower energy, the opposite peak
distribution is additionally penalized by $e^{-\Delta\Gamma}$, with
$\Delta\Gamma = \Gamma_Z - \Gamma_A$ on the order of eV, and then one would
expect to observe only one side of the distribution. These continuum-model
predictions turn out to be remarkably robust. To see this, below we present
the atomistic calculations of the relevant quantities, and proceed to simulate
example CNT type distributions.

\begin{figure}[t]
\includegraphics[width=\columnwidth]{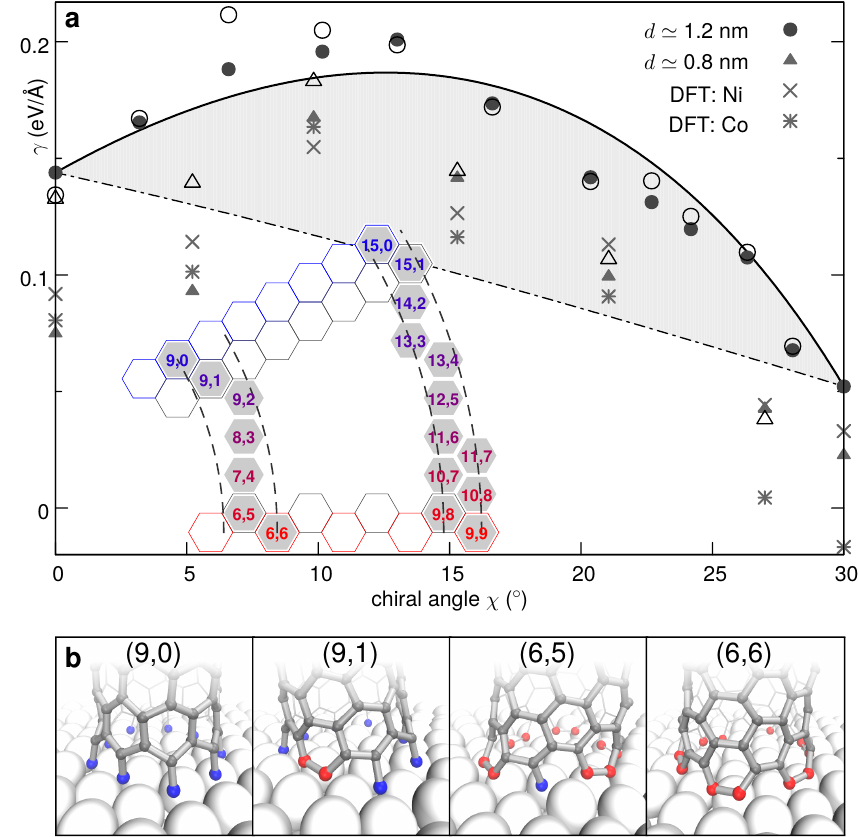}
\caption{\textbf{Chirality-dependent CNT--catalyst contact energies, governing
    the nucleation.} (\textbf{a}) Interface energies calculated with MD
  (circles and triangles) and fitted with analytical expression (solid line)
  for two CNT sets (inset; dashed arcs denote the range of diameter variation
  in each set). Static DFT calculations on Ni and Co are also shown. Open and
  filled symbols denote regular (hexagonal) and Klein $Z$ edge
  structures---see sample atomistic structures (\textbf{b}). The dash-dotted
  line corresponds to liquid catalyst case. In (\textbf{b}) the blue and red
  atoms highlight the $Z$- and $A$-edges, respectively.
\label{fig:2}}
\end{figure}

\textbf{Atomistic computations} were performed using a flat Ni$(111)$ slab to
represent the solid catalyst. We used a classical force-field MD sampling
complemented with static DFT computations. MD calculations were performed
using the canonical ($NVT$) ensemble with the ReaxFF force
field~\cite{vanduin01,mueller10} as implemented in the LAMMPS simulation
package~\cite{plimpton95,aktulga12}. DFT calculations were performed with the
local spin density approximation using the \textsc{Quantum ESPRESSO} package~\cite{giannozzi09}.

To investigate the chiral selectivity of \textbf{nucleation}, two sets of CNT
are chosen, with $d \approx 0.8$ and $1.2$~nm, to include the prominent in
experiments $(6,5)$ and $(9,8)$ CNT. \fig~\ref{fig:2}\textbf{a} shows the
calculated CNT--substrate interface energies. The atomistic structures for the
$(9,0)$, $(6,5)$, and $(6,6)$ CNT are shown in \fig~\ref{fig:2}\textbf{b},
where the tilting of the $(6,5)$ and $(9,1)$ CNT to reduce the interface
energy is seen clearly. We found that for the smaller-diameter set, for
near-$Z$ edges a Klein structure with dangling C atoms~\cite{klein94} is
favored over the standard closed-hexagons, whereas the larger-diameter set
shows little preference either way. All data display the same qualitative
behavior conforming to the above discussion, and are generally in good
quantitative agreement. Both bounds of the chiral angle range (achiral CNT)
are energy minima, and the energy is higher for $Z$ than for $A$ tubes. The
curves show the fit of $\gamma(\chi)$ using the analytical expression from
earlier work~\cite{liu10} for solid and liquid-like cases.

\begin{figure}[t]
  \includegraphics[width=\columnwidth]{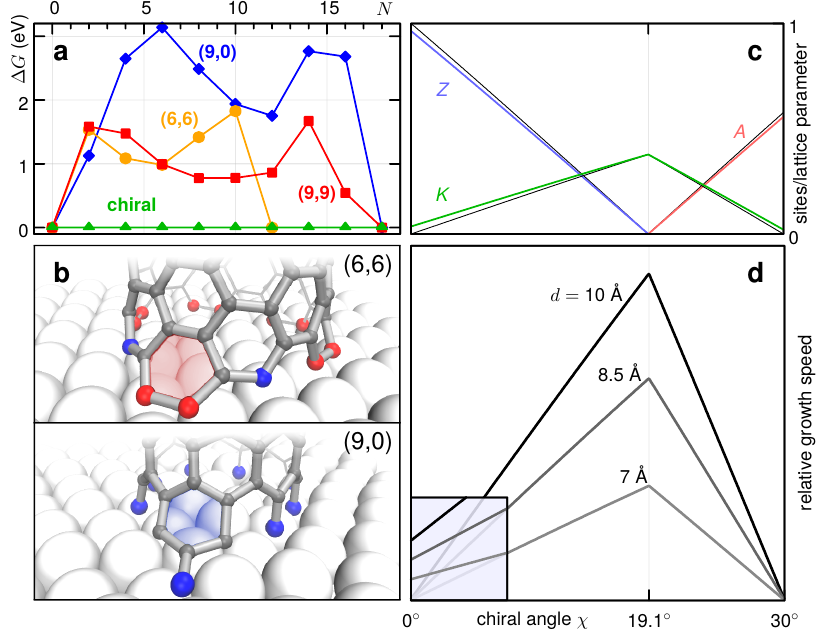}
  \caption{\textbf{Chirality-dependent growth rate of CNT.} (\textbf{a}) Free
    energy profiles during the growth of a new ring of hexagons on (red,
    orange) $A$ and (blue) $Z$ edges as a function of number of added atoms
    $N$. The green line corresponds to barrierless chiral edge
    growth. (\textbf{b}) The atomic configurations after first dimer addition,
    $N=2$. (\textbf{c}) Linear density of different site types on CNT edges as
    a function of chiral angle. (\textbf{d}) The resulting CNT growth rate as
    a function of chiral angle for several diameters (inset shows the effect
    of thermal kinks).
\label{fig:3}}
\end{figure}

Our computations pertaining to the growth kinetics are summarized in
\fig~\ref{fig:3}. We build upon our earlier approach for
graphene~\cite{artyukhov12}, adapting it for the case of
CNT. \fig~\ref{fig:3}\textbf{a} shows the energy changes with the addition of
a new row of carbon atoms, dimer by dimer, for $(6,6)$, $(9,9)$ and $(9,0)$
CNT. All three curves depart from the ``nucleation to kink-flow'' scenario of
graphene, the reason being the constantly changing tilt angle of the
CNT. However, both $A$ curves (red, orange) show essentially the same height
for the first dimer addition and the same maximum height (closer to the
end). The $Z$ curve bears the same qualitative character, having an initial
and a final maximum. The maximum height of each curve determines the free
energy barrier that needs to be overcome for successful addition of each new
row of hexagons, $\Delta G_A \approx 1.67$--1.86~eV and $\Delta G_Z > 3$~eV
(calculations for the $(9,0)$ CNT yield multiple intermediate structures with
topological defects of energies lower than perfect hexagonal structures, an
additional complication for $Z$-CNT growth). These are the terms that penalize
the pure $A$ and $Z$ tubes, compared to the chiral ones (green line in
\fig~\ref{fig:3}\textbf{a}), by an additional factor $\propto e^{-\Delta G}$
and thus effectively remove them from the product distribution, despite their
favorable cap-nucleation energies. The atomic configurations for the first
dimer addition to the $(6,6)$ and $(9,0)$ CNT are shown in
\fig~\ref{fig:3}\textbf{b}. \fig~\ref{fig:3}\textbf{c} shows the
concentrations of different site types---$Z$, $A$, and $K$ (kink)---as a
function of $\chi$. Thin black lines show the intrinsic, topologically
required values. Finally, \fig~\ref{fig:3}\textbf{d} shows the CNT growth
speed. Higher temperatures favor kink formation and promote the growth of
zigzag and armchair CNT. However, under realistic conditions, when $k_\text{B}
T\ll \Delta G$, the growth rate of achiral CNT is negligible. Among different
diameters $d$, the curvature of the wall penalizes insertion of C atoms into
small-diameter CNT, as in Eq.~\ref{eq:2}.

\begin{figure}[t]
  \includegraphics[width=\columnwidth]{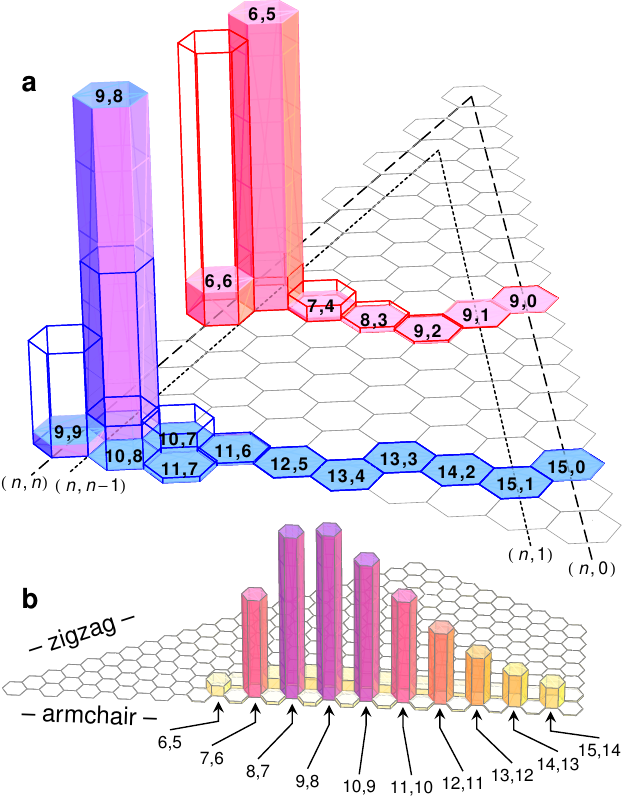}
  \caption{\textbf{Predicted CNT type distributions.} (\textbf{a})
    Distributions calculated directly based on MD computations for two CNT
    sets ($d\approx 0.8$~nm and 1.2~nm, each distribution is normalized
    separately). The empty bars show chirality distributions for liquid
    catalyst case. (\textbf{b}) Full $(n,m)$ distribution based on an
    analytical fit to MD interface energies. For all plots the temperature was
    artificially set to about three times the typical experimental value to
    make the heights visible.
\label{fig:4}}
\end{figure}

We now have all the ingredients to calculate the relative abundance of
different CNT types. The distributions for the two CNT data sets in
\fig~\ref{fig:4}\textbf{a} both show a strong predominance of $(n,n-1)$
near-armchair CNT. The selectivity of distributions is actually so strong that
one has to increase the temperature to $T=2700$~K when plotting these
distributions; at $T=900$~K, both show effectively single peaks for $(6,5)$ or
$(9,8)$ CNT. Further, one can use either data set to compute a general $(n,m)$
distribution through interface energy fitting. An example is shown in
\fig~\ref{fig:4}\textbf{b}. The peak in diameter distribution results from the
competition between the interface energy, favoring smaller $d$ in
Eq.~\ref{eq:1}, and a prefactor due to the configurational entropy of CNT
caps~\cite{penev14,brinkmann99}, favoring larger
diameters~\cite{dresselhaus96}, which scales approximately as $\sim
d^8$~\cite{reich05}. In a real CVD experiment, there will be additional
constraints on $d$, from the size of catalyst particles. Then the product
distribution will be a slice of \fig~\ref{fig:4}\textbf{b} with prominent
$(n,n-1)$ peaks, such as those shown in \fig~\ref{fig:4}\textbf{a}.

By similar logic, our theory suggests a possibility to highly selectively
achieve the near-$Z$ $(n,1)$ CNT, if a catalyst favors $Z$ interface over
$A$. While $(n,n-1)$ are always semiconducting, the $(n,1)$ series contains
all three CNT families (metallic and two semiconducting). Then, a control of
diameter would allow a selective synthesis of CNT of either conductivity
type. Moreover, when speculating on a possibility of catalyst-template exactly
matching a certain $(n,m)$ tube, we learn here that this would more likely
favor the \emph{one-index-off} tubes $(n,m\pm 1)$, to allow for rapid kinetics
at the cost of somewhat higher energy of the contact and nucleation.

We can also compare the simulated distributions to a liquid catalyst model
(dashed-dotted line in \fig~\ref{fig:2}\textbf{a} and the kink formation
energy $E_A=0$~\cite{ding09}). The results are shown in
\fig~\ref{fig:4}\textbf{a} as hollow bars and display much greater presence of
armchair CNT in the overall broader distribution. In reality, irregular and
highly mobile structure of liquid catalyst may flatten the energy landscape in
\fig~\ref{fig:2}\textbf{a} and thus further broaden the distribution. If
$E_A>0$, the fastest-growing tubes have $\chi =19.1^\circ$
(\fig~\ref{fig:3}\textbf{d}), which corresponds to $(2m,m)$ CNT. Finally, with
$\chi$-unbiased nucleation probability and $E_A \to 0$, one recovers the
proportionality result, $A\propto \chi$~\cite{ding09}.

In summary, the analysis above shows that the kinetic and thermodynamic
aspects of CNT growth must be considered concurrently. The growth kinetics is
aided by the kinks at the tube edge and thus favors the chiral types, in
proportion to their chiral angle. The thermodynamic nucleation barrier, on
solid catalyst, is lower for the kinkless edges of achiral tubes. In spite of
complex and random variability of numerous atomic structures in the process,
the overall product abundance can be summed up in a remarkably compact
mathematical expression: $x\, e^{-\beta\cdot x}$. For lower temperatures and
solid catalyst this function has a sharp maximum near zero, which explains the
observations of near-armchair nanotubes in experiments. Higher $T$ and liquid
catalyst make contact energies relatively equal ($\beta\to0$) and nucleation
of various types similarly probable, with the abundance then nearly
proportional to chiral angle. This demonstrates that the approach is
sufficiently comprehensive, being able to explain rather disparate facts
accumulated over decades of experiments, from broader chiral distributions to
very narrow, almost single type peaks. Furthermore, we believe that the gained
new insight must enable finding ways to engineer chiral-selective nanotube
production, thus advancing variety of long-awaited applications, all pending
availability of properly pure material.

\begin{acknowledgments}
\textbf{Acknowledgments:} Computer resources were provided by National Energy
Research Scientific Computing Center, which is supported by the Office of
Science of the U.S. Department of Energy under Contract No. DE-AC02-05CH11231;
XSEDE, which is supported by NSF grant OCI-1053575, under allocation
TG-DMR100029; and the DAVinCI cluster acquired with funds from NSF grant
OCI-0959097.
\end{acknowledgments}

\bibliographystyle{naturemag}
\bibliography{6598}

\end{document}